\documentclass[10pt,
               aps,
               pre,
               twocolumn,
               showpacs, 
               groupedaddress]{revtex4-2}
\usepackage[utf8]{inputenc}
\usepackage{graphicx} 

\usepackage{xcolor}
\graphicspath{{images/}}
\usepackage[caption=false]{subfig}
\usepackage{amsmath, amsfonts, amssymb}
\usepackage{hyperref}

\makeatletter
\newcommand*{\rom}[1]{\expandafter\romannumeral #1}
\makeatother

\begin{document}

\title{Inertial active Ornstein-Uhlenbeck particle in the presence of magnetic field}
\author{M Muhsin}
\affiliation{Department of Physics, University of Kerala, Kariavattom, Thiruvananthapuram-$695581$, India}

\author{M Sahoo}
\email{jolly.iopb@gmail.com}
\affiliation{Department of Physics, University of Kerala, Kariavattom, Thiruvananthapuram-$695581$, India}

\date{\today}

\begin{abstract}
\begin{description}
\item[Abstract] We consider an inertial active Ornstein-Uhlenbeck particle in an athermal bath. The particle is charged, constrained to move in a two-dimensional harmonic trap, and a magnetic field is applied perpendicular to the plane of motion. The steady state correlations and the mean square displacement are studied when the particle is confined as well as when it is set free from the trap. With the help of both numerical simulation and analytical calculations, we observe that inertia plays a crucial role in the dynamics in the presence of a magnetic field. In a highly viscous medium where the inertial effects are negligible, the magnetic field has no influence on the correlated behaviour of position as well as velocity. In the time asymptotic limit, the overall displacement of the confined harmonic particle gets enhanced by the presence of magnetic field and saturates for a stronger magnetic field. On the other hand, when the particle is set free, the overall displacement gets suppressed and approaches zero when strength of the field is very high. Interestingly, it is seen that in the time asymptotic limit, the confined harmonic particle behaves like a passive particle and becomes independent of the activity, especially in the presence of a very strong magnetic field. Similarly, for a free particle the mean square displacement in the long time limit becomes independent of activity even for a longer persistence of noise correlation in the dynamics.
\end{description}
\end{abstract}

\maketitle

\section{INTRODUCTION}
Recently, research on active matter has emerged as a vital area of research and attracted much attention in various fields of science\cite{bechinger2016active, Ramaswamy2017active, pietzonka2021oddity, magistris2015intro}. Active matter is a special class of nonequilibrium systems which is inherently or intrinsically driven far away from equilibrium. The particles in such a system are capable of self propelling by their own in the environment. They consume energy from the environment by means of their internal mechanisms and generate a spontaneous flow in the system \cite{magistris2015intro,dombrowski2004self}. These particles are termed as active or self-propelling particles. Examples of active matter include motile biological microorganisms like bacteria or unicellular protozoa \cite{berg1972chemo, machemer1972ciliary,corbyn2021stochastic}, artificially synthesized microswimmers like Janus particles \cite{walther2013janus, howse2007self}, microrobots, hexbugs\cite{scholz2018rotating}, etc. There exists some standard models like active Brownian particle (ABP) model to treat the dynamics of such particles at both single particle level as well as collective level \cite{hagen2009NonGaussian,hagen2011brownian, cates2013when,  kanaya2020steady, lowen2020inertial}. Recently, one of the simplest and nontrivial model known as active Ornstein-Uhlenbeck particle (AOUP) model \cite{lehle2018analyzing,  bonilla2019active, martin2021statistical} is proposed for modeling the over damped dynamics of such self-propelled particles. In ABP model, both the translational and rotational diffusion of the particles are taken into account while in AOUP model, the velocity of the particle follows the Ornstein-Uhlenbeck process. The AOUP model is explored in detail in literature as it makes the exact analytical calculations possible~\cite{szamel2014self,sandford2017pressure,marini2017pressure,Das2018confined,wittman2018effective,caprini2018linear,caprini2020time}. Both these models are successful in explaining many important features of active matter such as accumulation near boundary~\cite{marini2015towards,Gompper2020roadmap}, motility induced phase separation (MIPS)~\cite{cates2015motility} and so on. Unfortunately, inertia which is an important property of the physical systems, was not initially considered in these models. 

For macroscopic or massive self-propelling particles moving in a gaseous or low viscous medium, inertial effects become prominent and this poses some new challenges in the theoretical modeling of this kind of systems. Typically, millimeter sized particles while moving in a low viscous medium, are strongly influenced by inertia. Macroscopic swimmers\cite{gazzola2014scaling,saadat2017rules,gazzola2015gait} and flying insects\cite{sane2003aerodynamics} are some apt examples where inertia plays an important role in their dynamics, both at the single particle level as well as collective level \cite{lowen2020inertial}. Hence, inertia needs to be introduced in both AOUP as well as ABP models. Indeed, in some of the recent works, the introduction of inertia in these models could describe well the dynamics of active particles~\cite{caprini2021inertial, caprini2021spatial}. It is also reported that fine tuning of inertia results in some qualitative modification in the fundamental properties of active systems such as inertial delay between orientation dynamics and translational velocity of active particles\cite{scholz2018inertial}, development of different dynamical states\cite{dauchot2019dynamics}, motility induced phase separation\cite{mandal2019motility}, etc.

The stochastic dynamics of a charged particle in the presence of a magnetic field is an interesting problem with potential applications in plasma physics, astrophysics, electromagnetic theory, etc\cite{Singh1996stochastic, jayannavar1981orbital, saha2008nonequilibrium, aquino2009fluctuation, harko2016electro, lin2020seperation, jin2021collective}. According to Bohr-van Leeuwen (BvL) theorem\cite{nielsen1972niels, van1921problemes, dattagupta1997landau, Sahoo2007charged}, there is no orbital magnetism for a classical system of charged particles in equilibrium. However, when an inertial system exhibits activity in the dynamics, it does not follow the well known fluctuation dissipation theorem\cite{kubo1966fluct} and comes out of equilibrium. As a result, a nonzero orbital magnetism appears in the presence of magnetic field and the system passes through a magnetic phase transition depending on the complex interplay of the activity time and other time scales involved in the dynamics \cite{kumar2012classical, muhsin2021orbital}. 

When a time dependent magnetic field is applied to charged Brownian swimmers, it can either enhance or reduce the effective diffusion of swimmers\cite{sandoval2016magnetic}. On the other hand, the dynamics of a charged  active Brownian particle when subjected to a space dependent magnetic field, it induces inhomogeneity and flux in the system \cite{vuijk2020lorentz}. Similarly, under stochastic resetting, an active system in the presence of magnetic field yields some exotic steady state behaviour \cite{abdoli2021stochastic}. Motivated by these recent findings, herein, we explore the transport properties of a charged and inertial active Ornstein-Uhlenbeck particle in a viscous medium and under a static magnetic field. In particular, we show that inertia is necessary for the magnetic field to influence the dynamics.

The Brownian dynamics of an inertial charged particle in a magnetic field driven by an exponentially correlated noise and by a colored Gaussian thermal noise is already discussed in Refs.~\cite{Karmeshu1974brownian, paraan2008brownian, lisy2013brownian, baura2013study, lisy2014effect} and Ref.~\cite{das2017fokker}, respectively.
In these models, the dynamics is always mapped to it's thermal equilibrium limit, where the generalized fluctuation dissipation relation (GFDR) is satisfied. In our work, we consider the model as the dynamics of an active particle, which is different from the dynamics described in the previously discussed models in the sense that the active fluctuations are athermal and hence it can not be always mapped to an equilibrium limit. However, in the equilibrium limit of our model, where the fluctuation dissipation relation (FDR) is satisfied and in the vanishing limit of noise correlation time, some of our findings, especially the steady state diffusion shows similar behaviour as reported in Refs.~\cite{Karmeshu1974brownian,paraan2008brownian} for a free particle and in Refs.~\cite{lisy2013brownian,lisy2014effect} for a confined harmonic particle, respectively.

We have organized the paper in the following way. In Sec. II, we present our model, the methodology adopted, and introduction to the dynamical parameters of interest. The results and discussion are presented in Sec. III, followed by a summary in Sec. IV.

\section{MODEL AND METHOD}
We consider a charged active Ornstein-Uhlenbeck particle of mass $m$ self-propelling in a two dimensional (2D) plane. The particle is confined by a harmonic potential $U(x,y) = \frac{1}{2} k (x^2 + y^2)$ with $k$ as the harmonic constant. A magnetic field ${\bf B} = B {\bf \hat{z}}$ is applied perpendicular to the plane of the motion of particle, where ${\bf \hat{z}}$ is the unit vector along the Z-direction. The dynamics of the particle is given by Langevin's equation of motion \cite{arsha2021velocity, muhsin2021orbital, Sahoo2007charged}
\begin{equation}
m\ddot {\bf r}(t)=-\gamma {\bf v}(t) + \frac{|q|}{c}[{\bf v(t)}\times {\bf B}]-k{\bf{r}}(t)+\sqrt{2D}{\bf \xi}(t),
\label{eq:model-vector}
\end{equation}
where ${\bf \ddot{r}}=\dot{\bf v}$ is the acceleration of the particle and $m\ddot{\bf r}$ is the inertial force in the dynamics. The first term in the right hand side of Eq.~\eqref{eq:model-vector} is the viscous force on the particle because of the interaction of the particle with the surrounding medium, with $\gamma$ being the viscous coefficient of the medium. The second term represents the Lorentz force caused by the presence of magnetic field \cite{maxwell1873treatise} and the third term is the force exerted by the harmonic confinement. ${\bf \xi}(t)$ is the noise term which follows the Ornstein-Uhlenbeck process
\begin{equation}
t_c \dot{{\bf \xi}}(t) = -{\bf \xi}(t) + \eta(t),
\label{eq:noise-model}
\end{equation}
with $\eta(t)$ being the delta correlated white noise. $D$ is the strength of the Ornstein-Uhlenbeck noise \cite{sevilla2019generalized, woillez2020nonlocal, Das2018confined}. Further, $\bf{\xi}(t)$ satisfies the following properties
\begin{equation}
\langle \xi_\alpha(t) \rangle  = 0,  \qquad \langle \xi_\alpha(t) \xi_\beta(t^\prime) \rangle  = \frac{\delta_{\alpha\beta}}{2t_c}e^{\frac{-|t - t^\prime|}{t_c}}.
\label{eq:noise-stat}
\end{equation}
Here, $t_{c}$ is the noise correlation time or persistence time of the dynamics and $(\alpha, \beta) \in (X, Y)$. A finite correlation of noise for a time $t_{c}$ represents the persistence of activity upto $t=t_{c}$ and it decays exponentially with $t_{c}$. Hence, a finite and nonzero $t_{c}$ especially quantifies the activity of the system. In the $t_{c} \rightarrow  0$ limit, the active fluctuation becomes thermal and the system becomes passive in nature. In the present work, we consider $D=\gamma k_{B} T$ (fluctuation-dissipation relation or FDR) to have the typical thermal equilibrium limit of the dynamics at temperature $T$ \cite{fodor2016far,mandal2017entropy}. However, for a nonzero $t_{c}$, the dynamics is in nonequilibrium with an effective temperature which is different from the actual temperature of the system \cite{tailleur2009sedimentation}. In the long time limit, one can define this effective temperature with the self-propulsion speed of the active particle and can relate it to the strength of noise, $D$ \cite{ fily2012athermal}.

By defining $\Gamma = \frac{\gamma}{m},\ \omega_c = \frac{|q|B}{mc}, \ \text{and}\  \omega_0 = \sqrt{\frac{k}{m}}$ and introducing a complex variable $z(t) = x(t) + i\ y(t)$, Eq.~\eqref{eq:model-vector} can be rewritten in terms of $z(t)$ as
\begin{equation}
\ddot{z}(t) + \Gamma \dot{z}(t) - j \omega_c \dot{z}(t) + \omega_0^2 z(t) = \epsilon(t),
\label{eq:model-complex}
\end{equation}
where, $j =\sqrt{-1}$ and $\epsilon(t)=\frac{\sqrt{2 D}}{m}\left[ \xi_x(t) + j\ \xi_y(t)\right]$. By performing the Laplace transform of the complex variables $z(t)$ and $\dot{z}(t)$ $\left[ \mathcal{L}\{z\}(s) = \int\limits_0^\infty e^{-s t} z(t)\, dt\ \text{and} \ \mathcal{L}\{\dot{z}\}(s) = \int\limits_0^\infty e^{-s t} \dot{z}(t)\, dt \right]$, with initial conditions $z(0) = z_0$ and $\dot{z}(0) = v_0$, respectively and using the partial fraction method, the solution of the dynamics [Eq. \eqref{eq:model-complex}] can be obtained as
\begin{equation}
z(t) = \sum_{i=1}^2 b_i z_0 e^{s_i t} + \sum_{i=1}^2 a_i v_0 e^{s_i t} + \sum_{i=1}^2 a_i \int\limits_0^t e^{s_i (t - t^\prime)} \xi(t^\prime) dt^\prime.
\label{eq:model-solution} 
\end{equation}
Here $s_i$'s are given by
\begin{equation*}
s_{1,2} = \dfrac{-\Omega \pm \sqrt{\Omega^2 - 4\omega_0^2}}{2}, 
\label{eq:solution-si}
\end{equation*}
with $\Omega = \Gamma - j \omega_c$. The coefficients $a_i$'s and $b_i$'s are given by
\begin{equation*}
a_{1,2} = \pm \frac{1}{\sqrt{\Omega^2 - 4\omega_0^2}}\ \text{and} \  b_{1,2} = \pm \frac{\Omega \mp \sqrt{\Omega^2 - 4\omega_0^2}}{2\sqrt{\Omega^2 - 4\omega_0^2}},
\label{eq:solution-ai-bi}
\end{equation*}
respectively. In order to analyze the transport behaviour of such a system, we focus mainly on the mean displacement, steady state correlations, and mean square displacement. The mean displacement (MD), $\langle R(t) \rangle$ can be calculated from the relation
\begin{align}
\langle R(t) \rangle = & \langle z(t) - z(0) \rangle  \nonumber \\
 = &\langle x(t) - x(0) \rangle + j \langle y(t) - y(0) \rangle \nonumber \\ 
 = & a_1 \bigl[ e^{s_1 t} \left(s_1 z_0+v_0+\Omega  z_0\right) \nonumber \\
&-e^{s_2 t} \left(s_2
   z_0+v_0+\Omega  z_0\right) \bigr] - z_0.
\label{eq:solution-zavg}
\end{align}
The steady state position correlation [$C_r(t)$] and velocity correlation [$C_v(t)$] can be defined as
\begin{align}
C_r(t) & = \lim_{t^\prime \rightarrow \infty} \langle {\bf r}(t^\prime) \cdot {\bf r}(t^\prime + t) \rangle \nonumber \\
& = \lim_{t^\prime \rightarrow \infty} Re\left\{\langle z(t^\prime) z^*(t^\prime + t)  \rangle \right\}
\label{eq:def-cxt}
\end{align}
and
\begin{align}
C_v(t) & = \lim_{t^\prime \rightarrow \infty} \langle {\bf v}(t^\prime) \cdot {\bf v}(t^\prime + t) \rangle \nonumber \\
& = \lim_{t^\prime \rightarrow \infty} Re\left\{\langle \dot{z}(t^\prime) \dot{z}^*(t^\prime + t)  \rangle \right\}.
\label{eq:def-cvt}
\end{align}
In Eqs.~\eqref{eq:def-cxt} and \eqref{eq:def-cvt}, `$*$' denotes the complex conjugate and $Re\{\}$ represents the real part. Similarly, the mean square displacement(MSD), $\langle R^2(t) \rangle$ is given by the relation
\begin{align}
\langle R^2(t) \rangle & = \langle [{\bf r}(t) - {\bf r_0}]^2 \rangle \nonumber \\
& = \langle |z(t) - z_0|^2 \rangle.
\label{eq:def-msd}
\end{align}

\begin{figure}[!ht]
\includegraphics[scale=0.57]{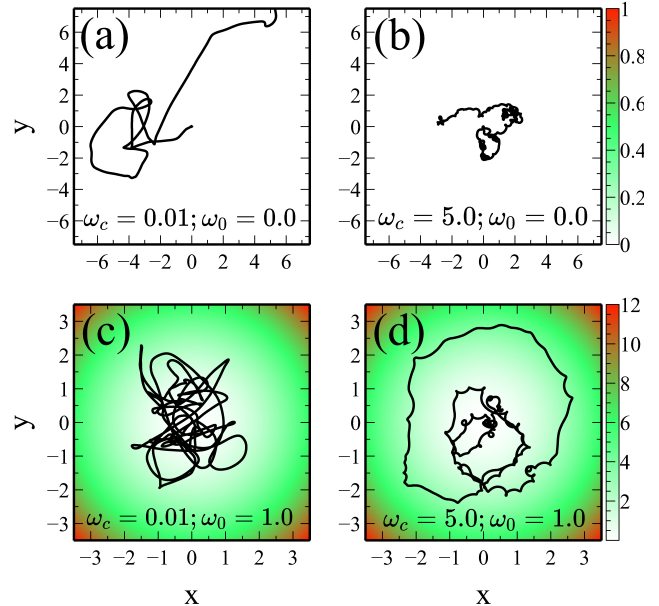}
\caption{Simulated trajectories of a free particle ($\omega_{0}=0$) in (a) and (b) and in the presence of harmonic confinement ($\omega_{0}=1.0$) in (c) and (d). The color map shows the strength of the harmonic confinement. Low magnetic field ($\omega_{c}=0.01$) is considered for (a) and (c) while high magnetic field ($\omega_{c}=5.0$) is taken for (b) and (d). The other common parameters are $m=1$, $\gamma=1.0$, and $t_{c}=1.0$.}
\label{fig:traj}
\end{figure}

The simulation of the dynamics [Eq.~\eqref{eq:model-vector}] is carried out using Heun's method algorithm \cite{gard1988intro} and Fox algorithm approaches\cite{fox1988fast}. A time step of $10^{-3}$ sec is chosen for each run of the simulation. For each realization, the simulation is run up to $10^5$ sec. The averages are taken over $10^5$ realizations after ignoring the initial transients (up to $10^3$ sec) in order for the system to reach the steady state. The detailed simulation results along with the analytical calculations are discussed in the following section.

\section{RESULTS AND DISCUSSION}
\begin{figure}[!ht]
    \centering
    \includegraphics[scale=0.42]{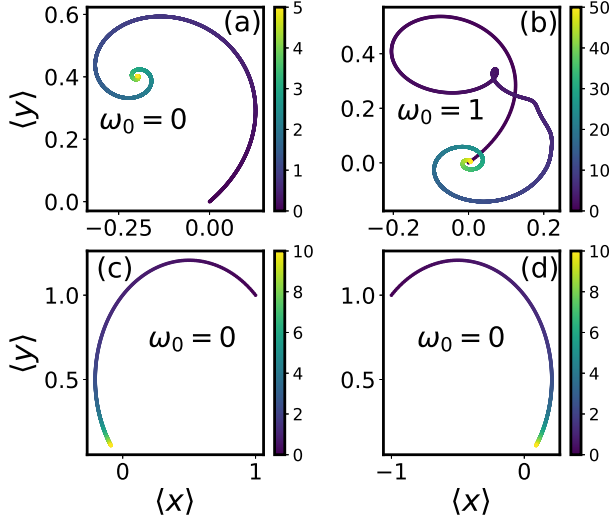}
    \caption{The 2D parametric plot of MD [Eq.~\eqref{eq:md-expansion}] is shown in (a), (c), and (d) for a free particle ($\omega_{0}=0.0$) and in (b) for a confined harmonic particle ($\omega_{0}=1.0$). The color map indicates the evolution of MD with time in both (a) and (b). For a free particle, MD attains a non-zero stationary value in the long time limit or at the steady state. The color map in (c) and (d) represents the evolution of this stationary MD with $\omega_{c}$ and $\gamma$, respectively. The other common parameters in (c) are: $z_0 = 0 + 0j,\ t_c = 1.0,\ v_0 = 1 + j,\ m = 1$, and $\gamma = 1$. Similarly, the other common parameters in (d) are: $z_0 = 0 + 0j,\ t_c = 1.0,\ v_0 = 1 + j,\ m = 1$, and $\omega_{c} = 1$.}
    \label{fig:md}
\end{figure}
In Fig.~\ref{fig:traj}, we have shown the simulated trajectories of the dynamics [Eq.~\eqref{eq:model-vector}] for a free particle [Fig.~\ref{fig:traj}(a) and (b)] as well as for a harmonically confined particle [Fig.~\ref{fig:traj}(c) and (d)]. The results presented in Figs.~\ref{fig:traj}(a) and (c) are for a low-strength magnetic field ($\omega_c=0.01$) whereas in Figs.~\ref{fig:traj}(b) and (d) are for a high-strength magnetic field ($\omega_c=5.0$). It is observed that in the absence of harmonic confinement ($\omega_{0}=0)$, the particle is set as free and the influence of a strong magnetic field makes the particle confined to a very small region. In this case, the directional movement of the self-propelling particle is dominated and the particle behaves as if it is trapped in the presence of a strong magnetic field [see Fig.~\ref{fig:traj}(b)]. On the other hand, when the particle is confined in a harmonic trap, it can not come out of the trap and under the influence of magnetic field, it precises around the field before coming back to the mean position in the long time limit. When the strength of the magnetic field is very large, the particle precises around the field for a longer time as well as travels a larger distance [see Fig.~\ref{fig:traj}(d)].

We have exactly calculated the mean displacement $\langle R(t) \rangle$ in the transient regime by expanding Eq.~\eqref{eq:solution-zavg} in the lower powers of $t$ as
\begin{equation}
    \langle R(t) \rangle = v_0t - \frac{1}{2} \left( v_0\Omega - z_0\omega_0^2 \right)t^2 + \mathcal{O}(t^3).
    \label{eq:md-expansion}
\end{equation}
The parametric plot of MD [$\langle y(t) \rangle$ vs $\langle x(t) \rangle$] is shown in Fig.~\ref{fig:md} when the particle is set free as well as when the particle is confined in a harmonic trap. The time asymptotic limit of the MD approaches zero value for a harmonically confined particle [$\lim\limits_{t\rightarrow \infty} \langle R(t) \rangle =0$] irrespective of the strength of the magnetic field. That is why in the long time limit, the particle reaches the center of harmonic trap ($z = 0$), which is nothing but the initial position ($z=0$) of the particle, as depicted in Fig.~\ref{fig:md}(b). However, this is not the case for a free particle [Fig.~\ref{fig:md}(a)]. For a free particle, it is found that $\lim\limits_{t\rightarrow \infty} \langle R(t) \rangle^{(f)} = \dfrac{v_0}{\Omega}$ and hence it depends on the magnetic field as well as on the viscosity of the medium. In the absence of magnetic field, i.e., for $\omega_c \rightarrow 0$ limit, $\lim\limits_{t\rightarrow \infty}\langle R(t) \rangle^{(f)} = \dfrac{v_0}{\Gamma}$,  which reflects that MD depends on the inertia of the particle. This is indeed consistent with the results reported in Ref.~\cite{nguyen_active_2022} for steady state MD. Figures~\ref{fig:md}(c) and (d) depict the 2D plots of the variation of steady state MD with $\omega_c$ and $\gamma$, respectively. It starts from the value $\dfrac{v_0}{\Gamma}$ for $\omega_c = 0$ and approaches zero value for a strong magnetic field [Fig.~\ref{fig:md}(c)]. Similarly, it starts from the value $\frac{jv_{0}}{\omega_{c}}$ for $\gamma=0$ and approaches zero for larger value of $\gamma$ [Fig.~\ref{fig:md}(d)]. This clearly indicates that the magnetic field has strong influence on MD only in the presence of inertia in the dynamics. In the absence of magnetic field, the MD increases with inertia and approaches zero for large $\gamma$ limit. It is also noteworthy that $\langle R(t) \rangle$ does not depend on the activity time or persistence time of the dynamics. This is because of the definition of statistical properties of the AOUP noise. In the lower time regime ($t\rightarrow 0$ limit), MD varies linearly with time and depends only on the initial velocity of the particle.

\begin{figure}[!ht]
\includegraphics[scale=0.31]{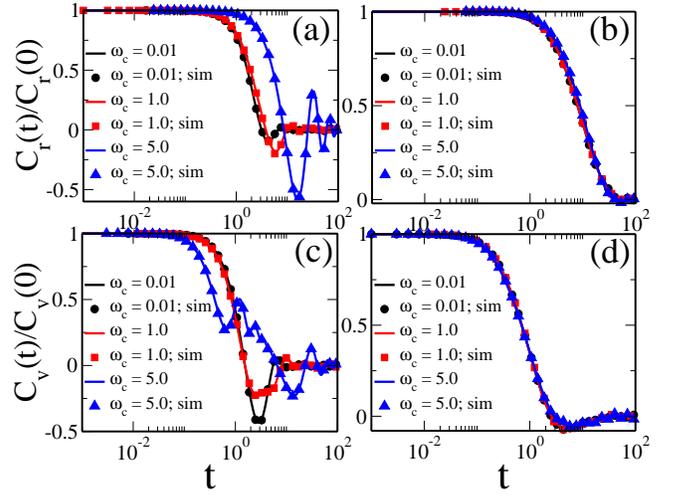}
\caption{Normalized $C_{r}(t)$ [Eq.~\eqref{eq:solution-cxt}] as a function of $t$ is shown in (a) and (b) and normalized $C_{v}(t)$ [Eq.~\eqref{eq:solution-cvt}] as a function of $t$ is shown in (c) and (d), respectively for a confined harmonic particle ($\omega_0=1$), obtained from the analytical calculations as well as from the simulation for different values of $\omega_c$. We have taken $\gamma = 1.0$  in (a) and (c) and $\gamma = 10.0$ in (b) and (d). The other common parameters are $\ t_c = 1$ and $\ m = 1$.}
\label{fig:cxt_cvt_wc}
\end{figure}
\begin{figure}[!hb]
\includegraphics[scale=0.31]{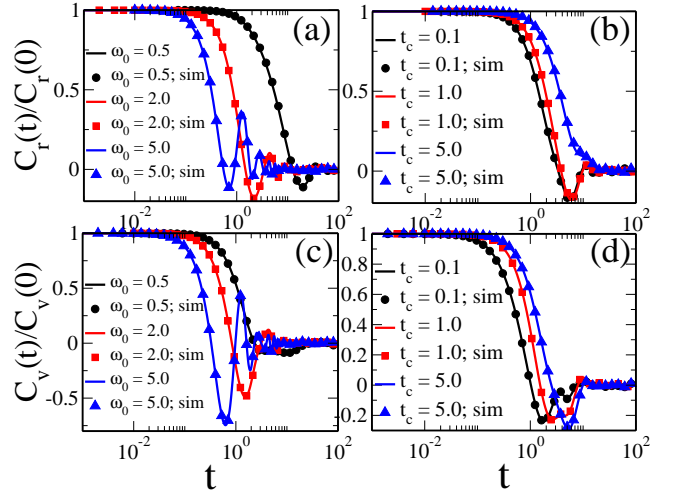}
\caption{Normalized $C_{r}(t)$ [Eq.~\eqref{eq:solution-cxt}] as a function of $t$ obtained from both analytical calculations and simulation for different values of $\omega_{0}$ and $t_c$ are shown in (a) and (b), respectively. Normalized $C_{v}(t)$ [Eq.~\eqref{eq:solution-cvt}] as a function of $t$ obtained from both analytical calculations and simulation for different values of $\omega_{0}$ and $t_c$ are shown in (c) and (d), respectively. We have taken $t_{c}=1$ in (a) and (c) and $\omega_0=1.0$ in (b) and (d). The other common parameters are $\omega_c = 1$, $m = 1$, and $\gamma = 1$.}
\label{fig:cxt_cvt_w0_tc}
\end{figure}

Next, we pay attention to the steady state behaviour of the position correlation $C_r(t)$ and velocity correlation $C_v(t)$. Substituting the solution $z(t)$ from Eq.~\eqref{eq:model-solution} and the noise properties from Eq.~\eqref{eq:noise-stat} in Eq.~\eqref{eq:def-cxt}, $C_r(t)$ can be calculated as
\begin{equation}
\begin{split}
	C_r(t) = Re\Bigg\{ & \sum_{i=1}^2 \sum_{j=1}^2 \frac{2a_i a_j^* D}{m^2} \Bigg[ \frac{t_c e^{-t/tc}}{(t_cs_i - 1) (t_cs_j^* + 1)} \\
	& - \frac{2 e^{s_j^* t}}{(s_i + s_j^*)(1 - t_c^2 s_j^{*2})} \Bigg] \Bigg\}.
\end{split}
\label{eq:solution-cxt}
\end{equation}
Similarly, substituting the solution $z(t)$ from Eq.~\eqref{eq:model-solution} and the noise properties from Eq.~\eqref{eq:noise-stat} in Eq.~\eqref{eq:def-cvt}, $C_v(t)$ can be calculated as
\begin{equation}
\begin{split}
	C_v(t) = Re\Biggl\{ & \sum_{i=1}^2 \sum_{j=1}^2 \frac{2c_i c_j^* D}{m^2} \Biggl[ \frac{t_c e^{-t/tc}}{(t_cs_i - 1) (t_cs_j^* + 1)} \\
	& - \frac{2 e^{s_j^* t}}{(s_i + s_j^*)(1 - t_c^2 s_j^{*2})} \Biggr] \Biggr\},
\end{split}
\label{eq:solution-cvt}
\end{equation}
where,
\begin{equation}
c_{1,2} = \frac{-\Omega \pm \sqrt{\Omega^2 - 4\omega_0^2}}{\sqrt{\Omega^2 - 4\omega_0^2}}.
\label{eq:solution-ci}
\end{equation}
For a confined harmonic particle, the normalized $C_r(t)$ and $C_v(t)$ are plotted as a function of $t$ in Fig.~\ref{fig:cxt_cvt_wc} for different values of $\omega_{c}$. The results presented in Figs.~\ref{fig:cxt_cvt_wc}(a) and (b) for $C_r(t)$ are for inertial ($\gamma=1$) and overdamped ($\gamma=10$) regimes, respectively. Similarly, the results presented in Figs.~\ref{fig:cxt_cvt_wc}(c) and (d) for $C_v(t)$ are for inertial ($\gamma=1$) and overdamped ($\gamma=10$) regimes, respectively. The obtained analytical results are in good agreement with the simulation. It is observed that with increase in magnetic field strength ($\omega_{c}$), the correlation in position persists for longer time before decaying to zero, whereas the velocity correlation decays faster with $\omega_{c}$ as expected. Most importantly, in the overdamped regime ($\gamma=10$), where the inertial effects are negligible, the magnetic field does not have influence on the correlated behaviour of either position or velocity.

The dependence of steady state correlations on harmonic confinement ($\omega_{0}$) and correlation time ($t_{c}$) are shown in Fig.~\ref{fig:cxt_cvt_w0_tc}. Both $C_{r}(t)$ and $C_v(t)$ decay faster with increase in $\omega_{0}$ whereas with increase in $t_{c}$, both quantities persist for longer time before decaying to zero.

Using Eq.~\eqref{eq:model-solution} in Eq.~\eqref{eq:def-msd}, the MSD of the harmonically confined particle $\langle R^2(t)\rangle$ can exactly be calculated as
\begin{widetext}
\begin{equation}
\begin{split}
    \langle R^2(t) \rangle = & \Biggl| a_1 \left(-\frac{z_0}{a_1}+\left(e^{s_1 t}-e^{s_2 t}\right) \left(v_0+\Omega  z_0\right)+s_1 z_0 e^{s_1 t}-s_2 z_0 e^{s_2 t}\right) \Biggr|^2 \\
   & +\sum_i\sum_j \frac{2a_i a_j^* D}{m^2} \Biggl[ \frac{t_c e^{t \left(s^*_j-\frac{1}{t_c}\right)}}{\left(t_c s_i+1\right) \left(1-t_c s^*_j\right)}+\frac{t_c e^{t
   \left(s_i-\frac{1}{t_c}\right)}}{\left(1-t_c s_i\right) \left(t_c s^*_j+1\right)} \\
   & +\frac{t_c \left(s_i+s^*_j\right)-2}{\left(s_i+s^*_j\right) \left(t_c s_i-1\right) \left(t_c s^*_j-1\right)}+\frac{e^{t \left(s_i+s^*_j\right)}
   \left(t_c \left(s_i+s^*_j\right)+2\right)}{\left(s_i+s^*_j\right) \left(t_c s_i+1\right) \left(t_c s^*_j+1\right)} \Biggr].
\end{split}
\label{eq:solution-msd}
\end{equation}
\end{widetext}
With the help of Taylor series expansion, Eq.~\eqref{eq:solution-msd} can be expanded in the powers of $t$ and by dropping the higher powers of $t$,  $\langle R^2(t) \rangle$ can be obtained as
\begin{equation}
\begin{split}
		\langle & R^2(t) \rangle = |v_0|^2t^2 - \frac{|v_0|^2 + 2(v_0^*z_0 + v_0z_0^*)\omega_0^2}{2}t^3 \\
		& + \frac{1}{12}\Biggl(\frac{6 D}{m^2 t_c} + \omega_0^2(5\Gamma-j\omega_c)(v_0z_0^* + z_0v_0^*) \\
		& + 3|z_0|^2\omega_0^4 + |v_0|^2(7\Gamma^2 -4\omega_0^2-\omega_c^2) \Biggr) t^4 +\mathcal{O}\left(t^5\right).
\end{split}
\label{eq:msd-shorttime}
\end{equation}

\begin{figure}[!hb]
\includegraphics[scale=0.52]{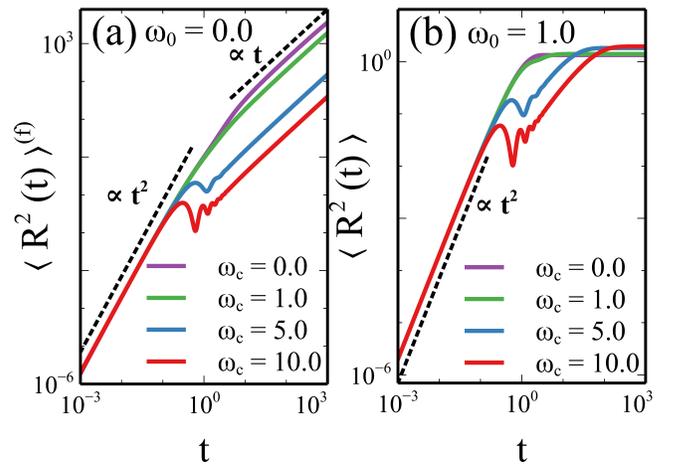}
\caption{MSD as a function of $t$ for different $\omega_c$ values (a) for a free particle [Eq.~\eqref{eq:solution-msd-free}] and (b) for a confined harmonic particle [Eq.~\eqref{eq:solution-msd}]. The other common parameters are $t_{c}=1$, $\gamma = 1$, and $m = 1$.}
\label{fig:msd_wc}
\end{figure}

We have plotted the MSD as a function of $t$ for a free particle as well as for a confined harmonic particle in Figs.~\ref{fig:msd_wc} (a) and (b), respectively for different values of $\omega_{c}$. From the exact calculation of MSD, it is confirmed that in the $t \rightarrow 0$ limit, $\langle R^2(t) \rangle$ is proportional to $t^{2}$, hence the dynamics is ballistic in nature. The initial ballistic regime ($\propto t^2$) depends solely on the initial velocity ($v_{0}$) of the article. When $v_0 = 0$, the initial regime of MSD is proportional to $t^4$. Dependence of MSD on $\omega_c$ appears starting from the fourth power of $t$. Since there is harmonic confinement, the particle cannot escape to infinity. Hence, in the long time regime, MSD attains a constant or saturated value $\langle R^2 \rangle_{st}$ [see Fig.~\ref{fig:msd_wc}(b)], which is given by the expression
\begin{equation}
\begin{split}
	\langle R^2 \rangle_{st} = & |z_0|^2 + \frac{2 D \left(t_c^2 \left(\omega _c^2+\Gamma ^2+\omega _0^2\right)\right)}{\Gamma  m^2 \omega _0^2
   \left(\left(t_c \left(\omega _0^2 t_c+\Gamma \right)+1\right)^2+t_c^2 \omega _c^2\right)} \\ & +\frac{2D\left(\Gamma  \omega _0^2 t_c^3+2 \Gamma  t_c+1\right)}{\Gamma  m^2 \omega _0^2
   \left(\left(t_c \left(\omega _0^2 t_c+\Gamma \right)+1\right)^2+t_c^2 \omega _c^2\right)}.
\end{split}
\label{eq:msd_st}
\end{equation}
This saturated value of MSD depends on $\omega_c$. In the $\omega_c \rightarrow 0$ limit, $\langle R^2 \rangle_{st}$ is obtained as
\begin{equation}
    \lim_{\omega_c \rightarrow 0} \langle R^2 \rangle_{st} = |z_0|^2 + \frac{2D(1 + t_c\Gamma)}{m^2 \omega_0^2 [\Gamma + t_c\Gamma(\Gamma + t_c \omega_0^2)]},
\end{equation}
which is same as that reported in Ref.~\cite{nguyen_active_2022} in the absence of magnetic field. In the presence of very strong magnetic field, i.e., in the $\omega_c\rightarrow\infty$ limit, the stationary MSD is simply $|z_0|^2 + \frac{2 D}{m^2 \Gamma \omega_0^2}$, which is independent of $t_c$. The same value of $\langle R^2 \rangle_{st}$ is obtained when we take the white noise limit, i.e., in the limit $t_c \rightarrow 0$. This confirms that the particle behaves like a passive particle in the presence of a high magnetic field. In thermal equilibrium limit of our model, MSD shows the similar behaviour as reported in Ref.~\cite{lisy2013brownian}. It is also observed that $\langle R^2 \rangle_{st}$ is an increasing function of $\omega_c$, and hence magnetic field enhances the overall displacement for a confined harmonic particle. This is very well reflected from Fig.~\ref{fig:msd_wc}(b).

The MSD for a free particle  $\langle R^2 (t) \rangle^{(f)}$ can be calculated by substituting $\omega_0=0$ and simplifying Eq.~\eqref{eq:solution-msd} as
\begin{widetext}
\begin{equation}
\begin{split}
    \langle R^2 (t)\rangle^{(f)} = & \frac{2 |v_0|^2 e^{-\Gamma t} \left(\cosh (\Gamma t)-\cos (\omega_c t )\right)}{\Gamma ^2+\omega_c^2} + \frac{2D}{m^2(\Gamma^2 + \omega_c^2)}\Biggl[ \\
    & \frac{e^{-\Gamma t} \left(2 \cos (\omega_c t) \left(t_c \omega_c^2 (\Gamma  t_c+1)+\Gamma  (\Gamma  t_c-2)
   (\Gamma  t_c-1)\right)-2 \omega_c \sin (\omega_c t) \left(t_c^2 \left(\Gamma ^2+\omega_c^2\right)-4 \Gamma
    t_c+2\right)\right)}{\left(\Gamma ^2+\omega_c^2\right) \left(t_c^2 \omega_c^2+(\Gamma  t_c-1)^2\right)} \\
    & + \frac{2 t_c^2 e^{-t \left(\Gamma +\frac{1}{t_c}\right)} \left(\omega_c \sin (\omega_c t) \left(t_c^2
   \left(\Gamma ^2+\omega_c^2\right)-2 \Gamma  t_c-1\right)+\cos (\omega_c t) \left(-\Gamma  \left(\Gamma ^2
   t_c^2-1\right)-t_c \omega_c^2 (\Gamma  t_c+2)\right)\right)}{\left(t_c^2 \omega_c^2+(\Gamma 
   t_c-1)^2\right) \left(t_c^2 \omega_c^2+(\Gamma  t_c+1)^2\right)} \\
   & -\frac{4 \Gamma }{\Gamma ^2+\omega_c^2}+\frac{2 t_c e^{-\frac{t}{t_c}} (\Gamma  t_c-1)}{t_c^2 \omega_c^2+(\Gamma  t_c-1)^2}+\frac{e^{-2 \Gamma  t} (\Gamma  t_c-1)}{\Gamma  \left(t_c^2 \omega_c^2+(\Gamma 
   t_c-1)^2\right)}+\frac{(\Gamma  t_c+1) (2 \Gamma  t_c+1)}{\Gamma  \left(t_c^2 \omega_c^2+(\Gamma 
   t_c+1)^2\right)} + 2t - 2t_c + 2t_c e^{\frac{-t}{t_c}} \Biggr].
\end{split}
    \label{eq:solution-msd-free}
\end{equation}
\end{widetext}
Expanding Eq.~\eqref{eq:solution-msd-free} in the powers of $t$, we get
\begin{equation}
\begin{split}
    \langle R^2 &(t) \rangle^{(f)} = |v_0|^2 t^2 - \Gamma   |v_0|^2 t^3 \\
    & + \left(\frac{D}{2 m^2 t_c}+\frac{7 \Gamma ^2 |v_0|^2}{12} - \frac{|v_0|^2
   \omega_c^2}{12}\right) t^4 +\mathcal{O}\left(t^5\right).
\end{split}
\label{eq:msd-shorttime-free}
\end{equation}
From this equation, it is confirmed that $\langle R^2 (t) \rangle^{(f)}$ depends on magnetic field and in the absence of magnetic field ($\omega_{c} \rightarrow 0$ limit), the result for $\langle R^2 (t) \rangle^{(f)}$ is consistent with that reported for a free particle in Ref.~\cite{nguyen_active_2022}. In the time asymptotic limit ($t \rightarrow \infty$), the MSD in Eq.~\eqref{eq:solution-msd-free} reduces to

\begin{equation}
\begin{split}
    \langle R^2 \rangle_{st}^{(f)} &= \frac{2D}{m^2(\Gamma^2 + \omega_c^2)}\Biggl( -\frac{4 \Gamma }{\Gamma ^2+\omega_c^2}+2 t -2 t_c \\
    & +\frac{(\Gamma  t_c+1) (2 \Gamma  t_c+1)}{\Gamma  \left(t_c^2 \omega_c^2+(\Gamma  t_c+1)^2\right)} \Biggr).
\end{split}
\label{eq:solution-msd-free-longtime}
\end{equation}

\begin{figure*}[!ht]
\includegraphics[scale=0.52]{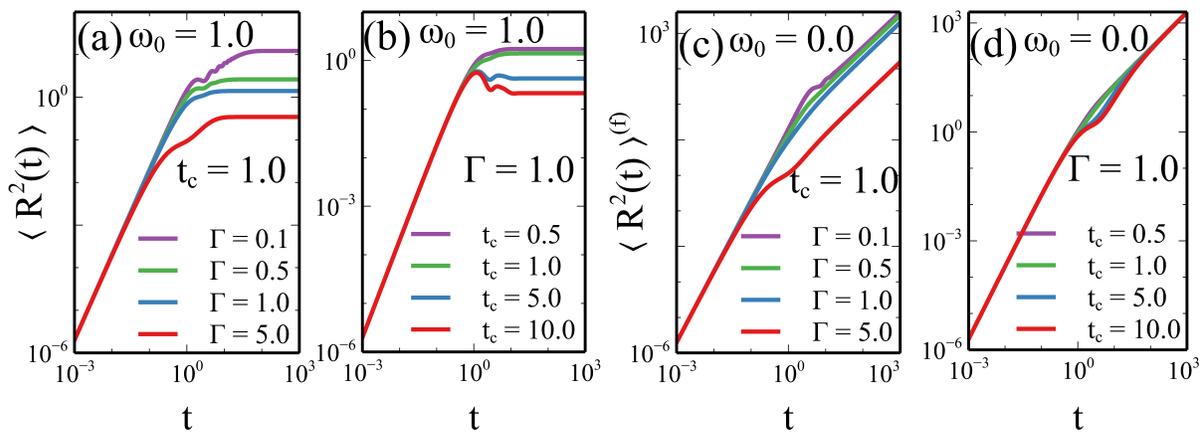}
\caption{For a harmonically confined particle ($\omega_{0}=1.0$), MSD as a function of $t$ [Eq.~\eqref{eq:solution-msd}] (a) for different $\Gamma$ values fixing $t_c=1.0$ and (b) for different $t_{c}$ values fixing $\Gamma=1.0$. For a free particle, MSD as a function of $t$ [Eq.~\eqref{eq:solution-msd-free}] (c) for different $\Gamma$ values fixing $t_c=1$ and (d) for different $t_c$ values fixing $\Gamma=1.0$. The common parameters are $\omega_c = 1.0$ and $m=1.0$.}
\label{fig:msd_G_tc}
\end{figure*}
Thus, the steady state MSD for a free particle depends on $\omega_{c}$ and approaches zero in $\omega_{c} \rightarrow \infty$ limit. This indicates that the presence of magnetic field suppresses the overall displacement of a free particle in contrast to that of a harmonically confined particle. These results are summarized in Fig.~\ref{fig:msd_wc}, where it can be seen that the initial ballistic regimes are similar for both the free and confined harmonic particle. However, in the long time regime, MSD is linearly proportional to $t$ for a free particle (diffusive in nature) but it approaches a stationary value for a confined harmonic particle (non-diffusive in nature). The steady state MSD for a free particle gets suppressed with magnetic field, whereas it gets enhanced for a confined harmonic particle. Other than these, we observe oscillations in the intermediate time regimes for both free and harmonically confined particle which could be due to the influence of magnetic field.
It is also to be noted that in $t_c\rightarrow \infty$ limit (with $t >> t_c$), $\langle R^2 \rangle_{st}^{(f)}$ can be obtained as
\begin{equation}
    \lim_{t_c\rightarrow\infty} \langle R^2 \rangle_{st}^{(f)} = \frac{2D}{m^2(\Gamma^2 + \omega_c^2)}\Biggl( -\frac{2 \Gamma }{\Gamma ^2+\omega_c^2}+2 t \Biggr),
\label{eq:solution-msd-free-hightc}
\end{equation}
which is independent of $t_c$. 

The MSD as a function of $t$ is plotted for different $\Gamma$ and $t_{c}$ values in Figs.~\ref{fig:msd_G_tc}(a) and (b) for a confined harmonic particle and in Figs.~\ref{fig:msd_G_tc}(c) and (d) for a free particle, respectively. It can be seen that for free particle in the time asymptotic limit, MSD is independent of $t_c$ while for confined harmonic particle, $t_c$ suppresses MSD. However, for both free and harmonically confined particle, the MSD gets suppressed with $\Gamma$. Since MSD for a free particle in the time asymptotic limit is proportional to $t$, the steady state diffusion coefficient for a free particle $\mathcal{D}_f$ can be calculated as
\begin{equation}
    \mathcal{D}_f = \lim_{t\rightarrow\infty} \frac{\langle R^2 (t)\rangle^{(f)}}{2t}= \frac{2D}{\gamma^2 + m^2\omega_c^2}.
\end{equation}
Substituting $\omega_{c}=\frac{qB}{mc}$ in the above equation, $\mathcal{D}_f$ can be simplified as
\begin{equation}
    \mathcal{D}_f = \frac{2D c^{2}}{\gamma^2 c^{2} + q^2 B^2}.
\end{equation}
Hence, $\mathcal{D}_f$ is independent of the mass of the particle but it depends on the magnetic field. It approaches zero when the particle is subjected to a strong magnetic field. In equilibrium limit ($D=\gamma k_{B} T$), the diffusive behaviour is found to be similar to that reported in Ref.~\cite{paraan2008brownian} and in the absence of magnetic field, the expression of $D_{f}$ is same as that reported in Ref.~\cite{nguyen_active_2022}.

\section{SUMMARY}
In this work, we have studied the motion of a charged inertial active Ornstein-Uhlenbeck particle in the presence of a magnetic field. One of the important observations is that, the magnetic field has strong influence in the dynamical behaviour of the particle because of the presence of inertia in the dynamics. The particle (if free) on an average covers a finite distance before settling down at a constant value in the long time limit. This constant value is found to be dependent on magnetic field which gets reduced with increase in field strength. On the contrary, if the particle is confined in a harmonic trap, it always comes back to the mean position of the trap, irrespective of the magnetic field. For a highly viscous medium, where the inertial influence is negligible, the dynamical behaviour of the particle is not affected by the magnetic field. Furthermore, the initial time regime of the mean square displacement is found to be similar and shows ballistic behaviour for both free and confined harmonic particle. On the other hand, the time asymptotic regime is diffusive for a free particle and non-diffusive for a harmonic particle. The ballistic regime for both free and confined harmonic particle gets reduced with increase in magnetic field strength.  

Surprisingly, for a harmonically confined particle, the steady state mean square displacement in the presence of a very strong magnetic field is same as that for a passive particle. When the strength of the magnetic field is very high, the steady state mean square displacement becomes independent of the field as well as on the noise correlation time or persistent time of the dynamics, ensuring the particle to behave like a passive particle. To understand this feature, it is further necessary to explore the relaxation behaviour of the dynamics and quantify the degree of irreversibility in terms of entropy production and non equilibrium temperature \cite{mandal2017entropy,fodor2016far}. Similarly, for a free particle, in the time asymptotic limit, the MSD becomes independent of activity despite the persistence of activity for a longer time. 

We believe that the results of our model are amenable for experimental verification and can be applied to implement the magnetic control on a charged active suspension by fine tuning the strength of the external magnetic field. It would be further interesting to explore the relaxation behaviour of the dynamics by introducing elasticity in the viscous solution \cite{igor2012visco, sevilla2019generalized}. Moreover, the inertial AOUP particle under the action of magnetic field  can be extended to more complex situation as in Ref.~\cite{vuijk2020lorentz, abdoli2021stochastic}.

\section{Acknowledgement}
M.S. acknowledges the start-up grant from UGC Faculty recharge program, Govt. of India for financial support.

%

%
\end{document}